\documentclass{article}

\usepackage{arxiv}
\usepackage{graphicx}
\usepackage[utf8]{inputenc} 
\usepackage[T1]{fontenc}    
\usepackage{hyperref}       
\usepackage{url}            
\usepackage{booktabs}       
\usepackage{amsfonts}       
\usepackage{nicefrac}       
\usepackage{microtype}      
\usepackage{lipsum}
\usepackage{color}

\usepackage{subfigure}
\usepackage{colortbl}
\usepackage{fontawesome}
\usepackage{multirow,tabularx}

\definecolor{lightgray}{gray}{0.9}

\newtheorem{example}{Example}

\newcommand{\remove}[1]{}

\newcommand{\chameleon}{\emph{chameleon} }

\title{The Chameleon Attack: Manipulating Content Display in Online Social Media}

\author{
  Aviad Elyashar, Sagi Uziel, Abigail Paradise, and Rami Puzis \\
  Telekom Innovation Laboratories and Department of Software and Information Systems Engineering,\\
  Ben-Gurion University of the Negev, Beer-Sheva, Israel \\
  \textit{\{aviade, sagiuz, abigailp\}@post.bgu.ac.il, puzis@bgu.ac.il} \\
}

\begin{document}

\maketitle

\begin{abstract}
Online social networks (OSNs) are ubiquitous attracting millions of users all over the world. 
Being a popular communication media OSNs are exploited in a variety of cyber attacks. 
In this article, we discuss the \chameleon attack technique, a new type of OSN-based trickery where malicious posts and profiles change the way they are displayed to OSN users to conceal themselves before the attack or avoid detection. 
Using this technique, adversaries can, for example, 
avoid censorship by concealing true content when it is about to be inspected; 
acquire social capital to promote new content while piggybacking a trending one; 
cause embarrassment and serious reputation damage by tricking a victim to like, retweet, or comment a message that he wouldn't normally do without any indication for the trickery within the OSN. 
An experiment performed with closed Facebook groups of sports fans shows that (1) \chameleon pages can pass by the moderation filters by changing the way their posts are displayed and (2) moderators do not distinguish between regular and \chameleon pages.   
We list the OSN weaknesses that facilitate the \chameleon attack and propose a set of mitigation guidelines. 
\end{abstract}


\section{Introduction}
\label{sec:introduction}

The following scenario is not a conventional introduction. Rather, it's a brief example to stress the importance and potential impact of the disclosed weakness, unless the countermeasures described in this article are applied. 
\begin{example}[A teaser]
\label{ex:teaser}
Imagine a controversial Facebook post shared by a friend of yours. 
You have a lot to say about the post, but you would rather discuss it in person to avoid unnecessary attention online. A few days later when you talk with your friend about the shared post, the friend does not understand what you're referring to.  
Both of you scan through his/her timeline and nothing looks like that post.
The next day you open Facebook and discover that in the last six months you have joined three Facebook groups of Satanists; you actively posted on a page supporting an extreme political group (although your posts are not directly related to the topics discussed there), and you liked several websites leading to video clips with child abuse. 
A terrible situation that could hurt your good name especially if you are a respected government employee!
\end{example}

At the time of submission of this article, the nightmare described in Example~\ref{ex:teaser} is still possible in major online social networks (OSNs) (see Section~\ref{sec:susceptibility}) due to a conceptual design flaw. 

Today, OSNs are an integral part of our lives~\cite{boyd2007social}. 
They are powerful tools for disseminating, sharing and consuming information, opinions, and news~\cite{kwak2010twitter}; and for expanding connections \cite{gilbert2009predicting}, etc. 
However, OSNs are also constantly abused by cybercriminals who exploit them for malicious purposes including spam and malware distribution \cite{lee2010uncovering}, harvesting personal information \cite{boshmaf2011socialbot}, infiltration  \cite{elyashar2013homing}, and spreading misinformation \cite{ferrara2015manipulation}.
Bots, fake profiles, and fake information are all well-known scourges being tackled by OSN providers, academic researchers, and organizations around the world with various levels of success. 
It is extremely important to constantly maintain the content of social platforms and service-wise, in order to limit abuse as much as possible.

In order to provide the best possible service to their users, OSNs allow users to edit or delete  published content~\cite{facebook_help_edit_post}, edit user profiles, and update previews of linked resources, etc. 
These features are important in order to keep social content up to date, to correct grammatical or factual errors in published content, and eliminate abusive content. 
Unfortunately, they also open an opportunity for a scam where OSN users are tricked into engaging with seemingly appealing content that is later modified. 
This type of scam is trivial to execute and is out of the scope of this article.

Facebook partially mitigates the problem of modifications made to posts after their publication by displaying an indication that a post was edited. 
Other OSNs, such as Twitter or Instagram, do not allow published posts to be edited. 
Nevertheless, the major OSNs (Facebook, Twitter, and LinkedIn) allow publishing redirect links, and they support link preview updates. 
This allows changing the way a post is displayed without any indication that the target content of the URLs has been changed.  

In this article, we present a novel type of OSN attack termed the \chameleon attack, where the content (or the way it is displayed) is modified over time in order to create social traction before executing the attack (see Section~\ref{sec:chameleon_attack}).
We discuss the OSN misuse cases stemming from this attack and their potential impacts in Section~\ref{sec:misuse}. 
We review the susceptibility of seven major OSN platforms to the \chameleon attack in Section~\ref{sec:susceptibility} and present the results of an intrusion into closed Facebook groups facilitated by it in Section~\ref{sec:group_infiltration_experiments}.
A set of suggested countermeasures that should be applied to reduce the impact of similar attacks in the future is suggested in  Section~\ref{sec:mitigation}. 

The contribution of this study is three-fold:
\begin{itemize}
    \item We present a new OSN attack termed the \chameleon attack including an end-to-end demonstration on major OSNs (Facebook, Twitter, and LinkedIn).
    \item We present a social experiment on Facebook showing that chameleons facilitate infiltration into closed communities.
    \item We discuss multiple misuse cases and mitigation methods from which we derive a recommended course of action to OSNs.   
\end{itemize}

\section{Background on redirection and link preview}

\paragraph{Redirection}
Redirection is a common practice on the Web that helps Internet users to find relocated resources, to use multiple aliases for the same resource, and to shorten long and cumbersome URLs.  
As a case in point, the use of URL shortening services is very common within OSNs.

There are two types of redirect links: server, and client redirects. 
In case of a server-side redirect, the server returns the HTTP status code 301 (redirect) with a new URL. 
Major OSNs follow server-side redirects up to the final destination in order to provide their users with a preview of the linked Web resource.  
In the case of a client-side redirect, the navigation process is carried out by a JavaScript command executed in the client's browser.
Since the OSNs do not render the Web pages they do not follow the client redirects up to the final destination.  

\paragraph{Short links and brand management}
There are many link redirection services across the Web that use 301 server redirects for brand management, URL shortening, click counts and various website access statistics. 
Some of these services that focus on brand management, such as \url{rebrandly.com}, allow their clients to change the target URL while maintaining the aliases. 
Some services, such as \url{bitly.com}, require a premium subscription to change the target URL. 
The ability to change the target URL without changing the short alias is important when businesses restructure their websites or move them to a different web host. 
Yet, as will be discussed in Section~\ref{sec:chameleon_attack}, this feature may be exploited to facilitate the \chameleon attack.  

\paragraph{DNS updates}
DNS is used to resolve the IP address of a server given a domain name. 
The owner of the domain name may designate any target IP address for his/her domain and change it at will. 
The update process may take up to 24 hours to propagate. 
Rapid DNS update queries, known as Fast Flux, are used by adversaries to launch spam and phishing campaigns. 
Race conditions due to the propagation of DNS updates cause a domain name to be associated with multiple, constantly changing IP addresses at the same time.

\paragraph{Link previews}
Generating and displaying link previews is an important OSN feature that streamlines the social interaction within the OSN.
It allows the users to quickly get a first impression of a post or a profile without extra clicks.   
Based on the met-tags of the target page, the link preview, usually includes a title, a thumbnail, and a short description of the resource targeted by the URL~\cite{kopetzky1999visual}.

When shortened URLs or other server-side redirects are used, the OSN follows the redirection path to generate a preview of the final destination. 
These previews are cached due to performance considerations.  
The major OSNs update the link previews upon request (see Section~\ref{sec:weaknesses} for details). 
In the case of client-redirect, some OSNs (e.g. Twitter) use the meta-tags of the first HTML page in the redirect chain. 
Other, e.g. Facebook, follow the client redirect up to the final destination. 

\section{The Chameleon Attack}
\label{sec:chameleon_attack}

The \chameleon attack takes advantage of link previews and redirected links to modify the way that published content is displayed within the OSN without any indication for the modifications made.
As part of this attack, the adversary circumvents the content editing restrictions of an OSN by using redirect links.

\begin{figure}[h!]
\centering
\includegraphics[scale=0.5]{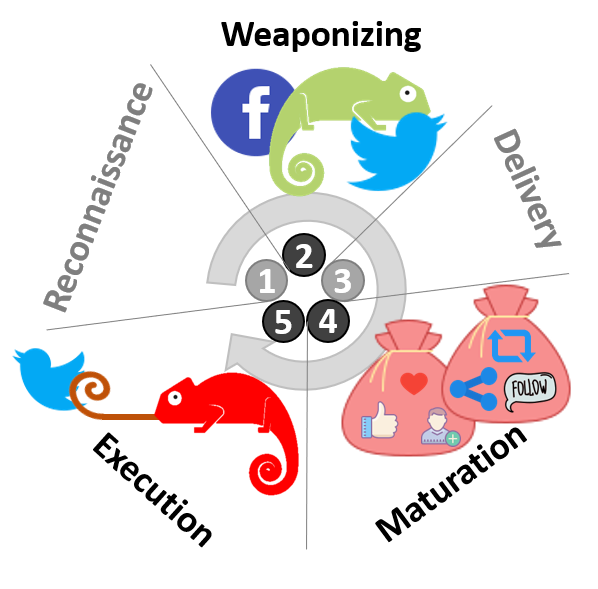}
\caption{\label{fig:chameleon_attack_phases}Chameleon Attack Phases.}
\end{figure}
\footnotetext{The authors would like to thank the icons website (\url{https://icons8.com)}}

We align the phases of a typical \chameleon attack to a standard cyber kill chain as follows: 
\begin{enumerate}
\item\textbf{Reconnaissance} (out of scope):
The attacker collects information about the victim using standard techniques to create an appealing disguise for the \chameleon posts and profiles. 
 
\item\textbf{Weaponizing} (main phase):
The attacker creates one or more redirection chains to Web resources (see Required Resources in Section~\ref{sec:resources}).  
The attacker creates \chameleon posts or profiles that contain the redirect links.

\item\textbf{Delivery} (out of scope):
The attacker attracts the victim's attention to the \chameleon posts and profiles, similar to phishing or spear-phishing attacks. 
 
\item\textbf{Maturation} (main phase):
The \chameleon content builds trust within the OSN, collects social capital, and interacts with the victims.  
This phase is inherent to spam and phishing attacks that employ fake OSN profiles. 
But since such attacks are not considered as sophisticated and targeted, this phase is typically ignored in standard cyber kill chains or is referred to by the general term of  \emph{social engineering}. 
Nevertheless, building trust within an OSN is very important for the success of both targeted and un-targeted \chameleon attacks.  

\item\textbf{Execution} (main phase):
The attacker modifies the display of the \chameleon posts or profiles by changing the redirect target and refreshing the cached link previews. 
\end{enumerate}
Since the \chameleon attack is executed outside the victim's premises there are no lateral movement or privilege escalation cycles. 
This attack can be used during the reconnaissance phase of a larger attack campaign or to reduce the cost of weaponizing any OSN based attack campaign (see examples in Section~\ref{sec:misuse}). 
The most important phases in the execution flow of the \chameleon attack are \emph{weaponizing}, \emph{maturation}, and \emph{execution} as depicted in  Figure~\ref{fig:chameleon_attack_phases}. 
The attacker may proceed with additional follow-up activities depending on the actual attack goal as described in Section~\ref{sec:misuse}.

\begin{figure*}
  \centering
    \includegraphics[width=0.16\linewidth]{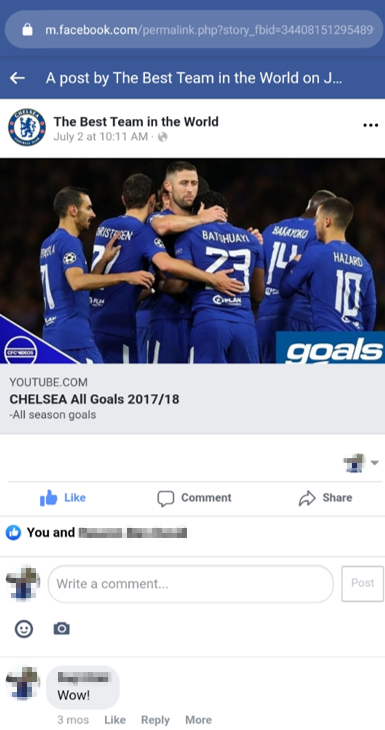}
    \includegraphics[width=0.16\linewidth]{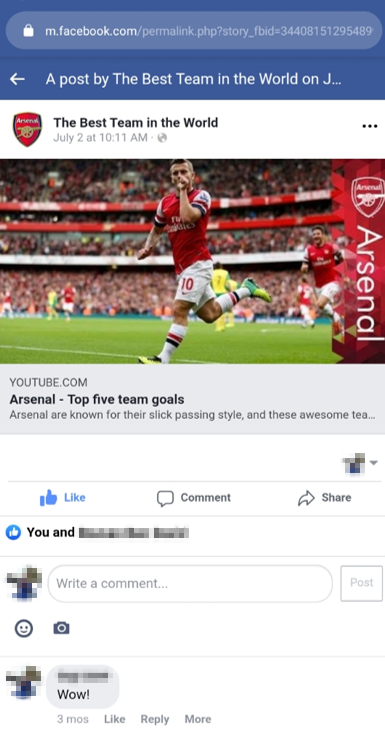}
    \includegraphics[width=0.16\linewidth]{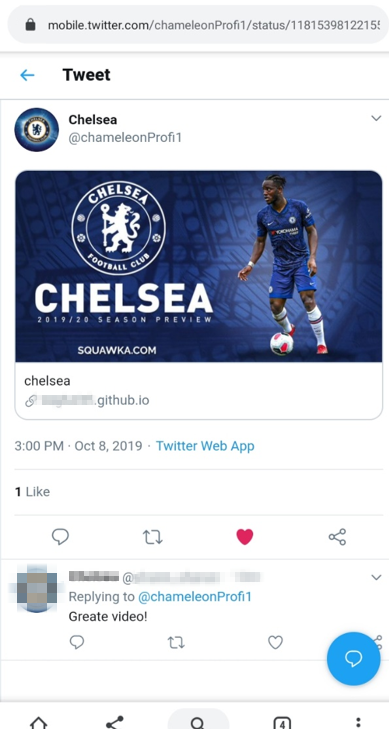}
    \includegraphics[width=0.16\linewidth]{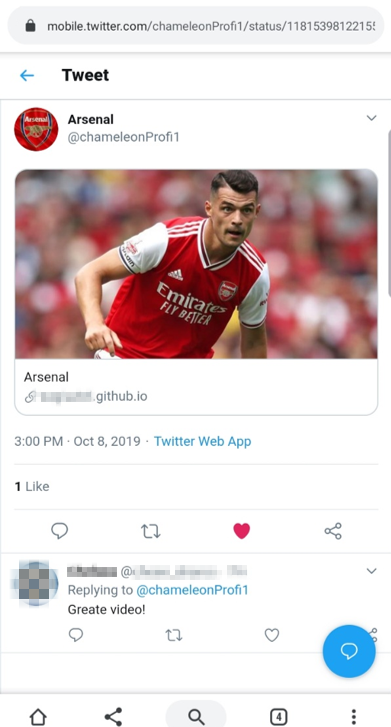}
    \includegraphics[width=0.16\linewidth]{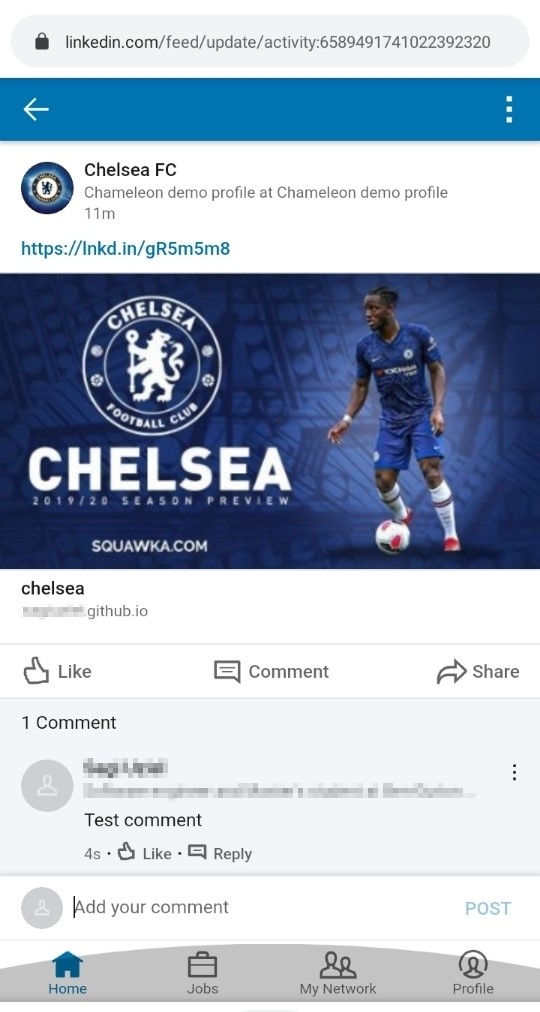}
    \includegraphics[width=0.16\linewidth]{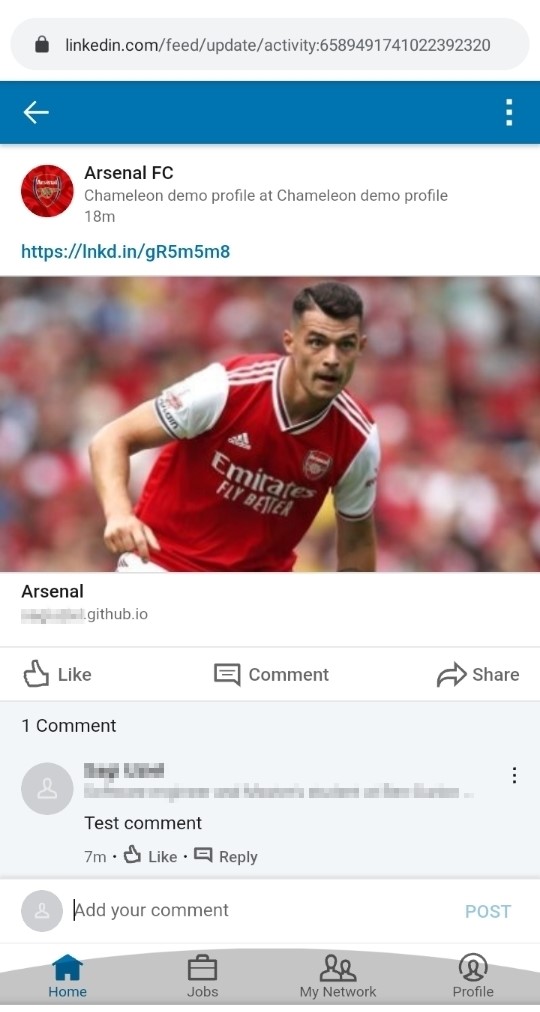}
  \caption{Changing post display in Facebook, Twitter, and LinkedIn (from left to right)}
  \label{fig:posts}
\end{figure*}

\begin{figure*}
  \centering
    \includegraphics[width=0.16\linewidth]{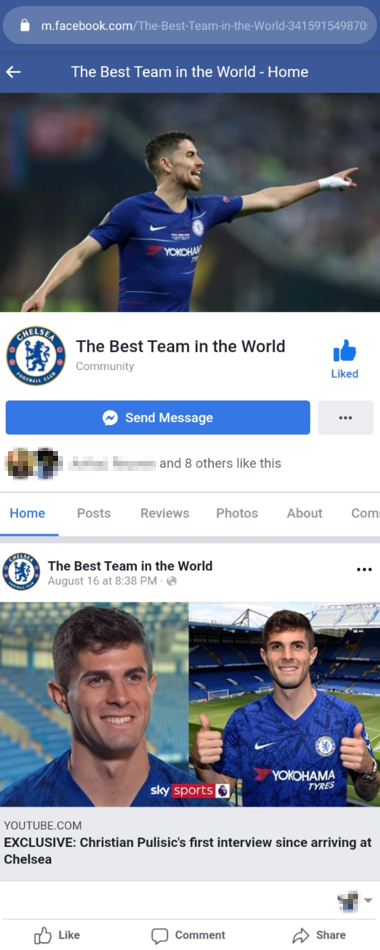}
    \includegraphics[width=0.16\linewidth]{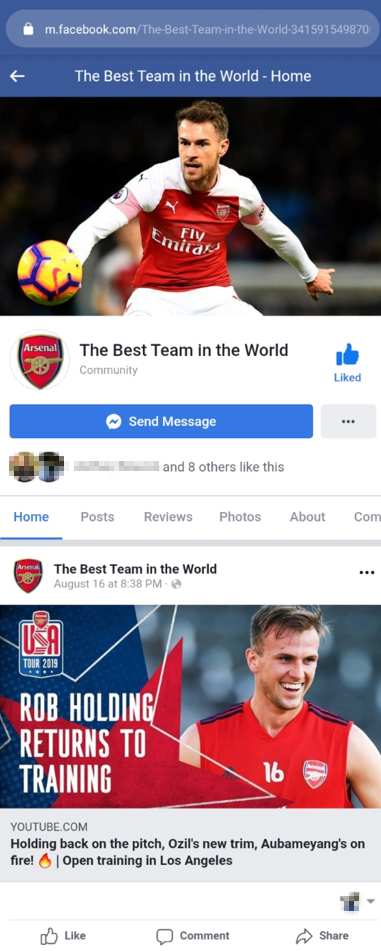}
    \includegraphics[width=0.16\linewidth]{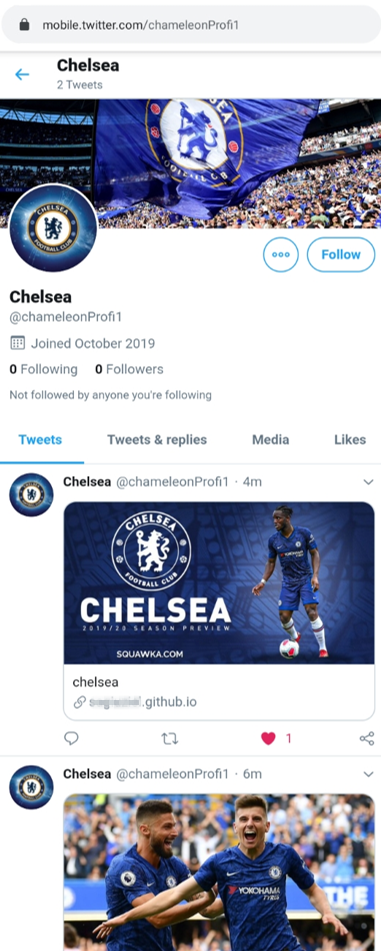}
    \includegraphics[width=0.16\linewidth]{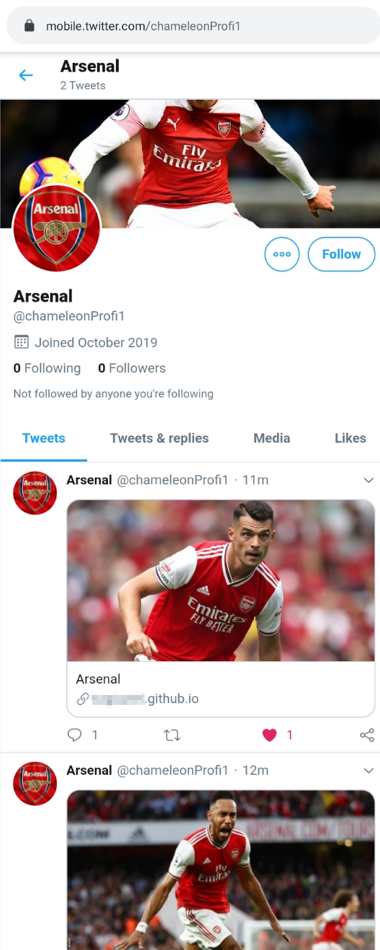}
    \includegraphics[width=0.16\linewidth]{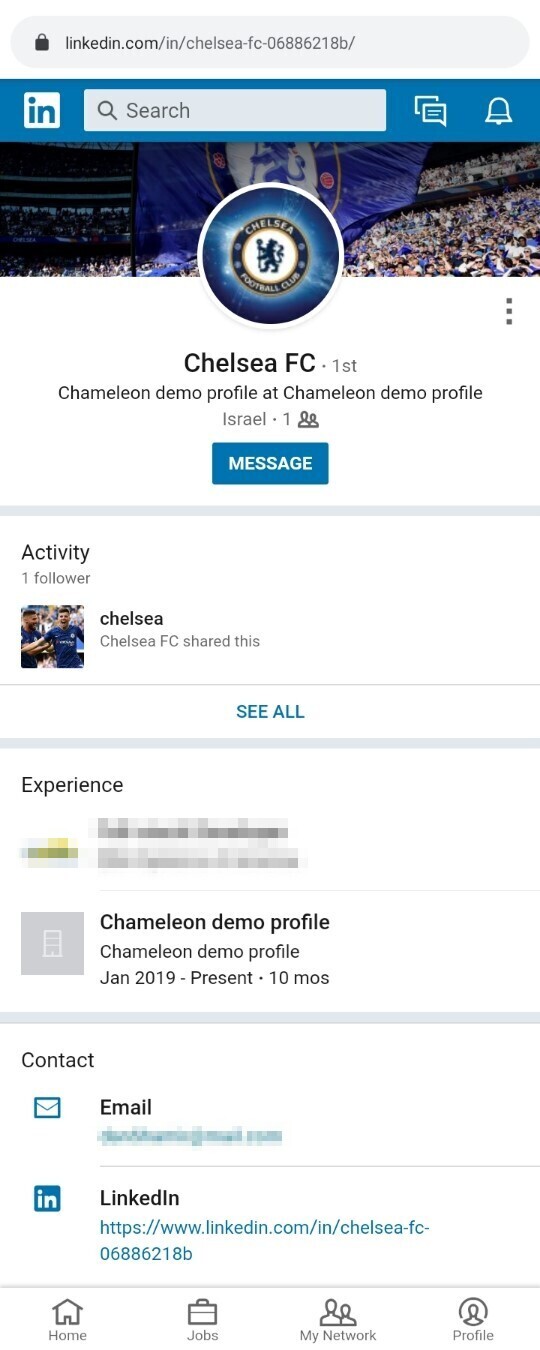}
    \includegraphics[width=0.16\linewidth]{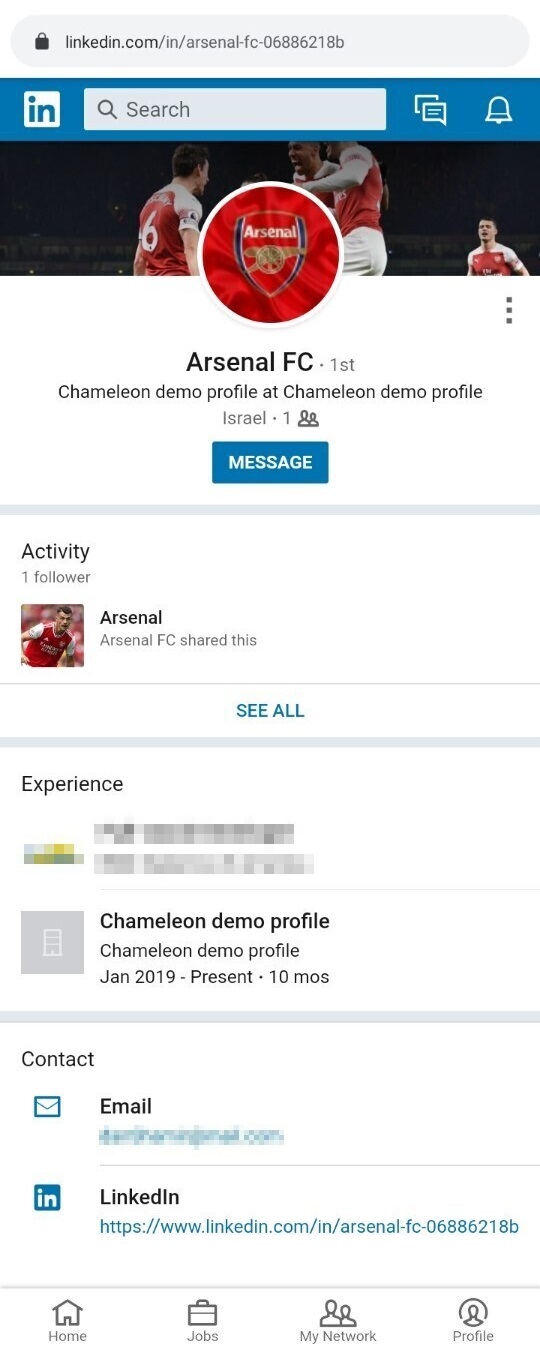}
  \caption{Changing page display in Facebook, Twitter, and LinkedIn (from left to right)}
  \label{fig:pages}
\end{figure*}

\subsection{A Brief Showcase}
\label{sec:showcase}
In order to demonstrate how a \chameleon attack looks from the user's perspective, we show here examples of \chameleon posts and
\chameleon profiles.\footnote{A demo  \chameleon post is available at \url{https://www.facebook.com/permalink.php?story_fbid=101149887975595&id=101089594648291&__tn__=-R}}  
 
The link preview in this post will change each time you click the video. 
It may take about 20 seconds and requires refreshing the page.   

\paragraph{Chameleon Post} 
Figures~\ref{fig:posts} (1,2) present the same post on Facebook with two different link previews.  
Both versions of the post lead to YouTube.com and are displayed accordingly. 
There is no indication of any modification made to the post in either of its versions because the actual post was not modified. 
Neither is there an edit history, for the same reason.
Likes and comments are retained. 
If the post was shared, the shares will show the old link preview even after it was modified in the original post. 

Similarly, Figure~\ref{fig:posts} (3,4) and (5,6) present two versions of the same post on Twitter and LinkedIn respectively. 
There is no edit indication nor edit history because Twitter tweets cannot be edited.  
As with Facebook, likes, comments, and retweets are retained after changing the posted video and updating the link preview.  
Unlike Facebook, however the link previews of all retweets and all LinkedIn posts that contain the link will change simultaneously.

\paragraph{Chameleon Profile} 
Figure~\ref{fig:pages} presents example of a \chameleon page on Facebook and a \chameleon profile on Twitter. 
Since the technique used to build \chameleon profiles and \chameleon pages are similar as well as their look and feel, in the rest of this paper we will use the terms pages and profiles interchangeably. 
All OSNs allow changing the background picture and the description of profiles, groups, and pages. 
A \chameleon profile is different from a regular profile by \chameleon posts included alongside neutral or personal posts. 
This way a Chelsea fan (Figure~\ref{fig:pages}.1) can pretend to be an Arsenal fan (Figure~\ref{fig:pages}.2) and vice versa.

\subsection{Required Resources}
\label{sec:resources}
The most important infrastructure element used to execute the \chameleon attack is a redirection service that allows the attacker to modify the redirect target without changing the alias. 
This can be implemented using a link redirection service or a website controlled by the adversary. 
For the attack to succeed in the former case, the link redirection service must allow modifying the target link for a previously defined alias.  
This is the preferred infrastructure to launch the \chameleon attack. 

In the latter case, if the attacker has control over the redirection server then a server-side 301 redirect can be used, seamlessly utilizing the link preview feature of major OSNs.
If the attacker has no control over the web server, he/she may still use a client-side redirect. 
He/she will have to supply the required metadata for the OSN to create link previews. 

If the attacker owns the domain name used to post the links, he/she may re-target the IP address associated with the domain name to a different web resource. 
Fast flux attack infrastructure can also be used; however, this is overkill for the \chameleon attack and may cause the attack to be detected~\cite{holz2008measuring}.

\subsection{Example Instances}
\label{sec:misuse}
In this section, we detail several examples of misuse cases~\cite{mcdermott2000eliciting} which extend the general \chameleon attack. 
Each misuse case provides a specific flavor of the attack execution flow, as well as the possible impact of the attack.

\subsubsection{Incrimination and Shaming} 
This flavor of the \chameleon attack targets specific users. 
Shaming is one of the major threats in OSNs \cite{goldman2015trending}. 
Its impact can be greatly amplified if the adversary employs chameleons and the victim is careless enough to interact with content posted by a dubious profile or page.  

\paragraph{Execution flow}
The attacker performs the (1) \emph{reconnaissance} and (3) \emph{delivery} phases using standard techniques, similar to a spear-phishing attack.\footnote{Here and in the rest of this section, numbers in parentheses indicate the attack phases in the order they are performed in each misuse case.}

\noindent(2) During the \emph{weaponizing} phase, the attacker creates \chameleon posts that endorse a topic favored by the victim. 
For example, he/she may post some new clips of music by the victim's favorite band. 
Each post includes a redirect link that points to an appropriate YouTube video or similar Web resource, but the redirection is controlled by the attacker. 

\noindent(4) During the \emph{maturation} phase, the victim shows their appreciation of seemingly appealing content by following the \chameleon page, linking, retweeting, commenting, or otherwise interacting with the \chameleon posts. 
Unlike in spear-phishing where the victim is directed to an external resource or is required to expose his/her personal information, here standard interactions that are considered safe within OSNs are sufficient to affiliate the victim with the \chameleon posts. 
This significantly lowers the attack barrier.

\noindent(5) Finally, immediately after the victim's interaction with the \chameleon posts, the adversary switches their display to content that opposes the victim's agenda in order to cause maximal embarrassment or political damage. 
The new link preview will appear in the victim's timeline. 
The OSN will amplify this attack by notifying the victim's friends (Facebook) and followers (Twitter) about the offensive posts liked, commented, or retweeted by the victim.

\paragraph{Potential impact}
At the very least, such an attack can cause discomfort to the victim. 
It can be life-threatening in cases when the victim is a teenager.
And it can have far-reaching consequences if used during political campaigns.

\subsubsection{Long Term Avatar Fleet Management} 
Adversaries maintain fleets of fake OSN profiles termed avatars in order to collect intelligence, infiltrate organizations, disseminate misinformation, etc.  
In order to avoid detection by machine learning algorithms and build long term trust within the OSN sophisticated avatars need to be operated by a human \cite{elyashar2016guided,paradise2017creation}. 
The maturation process of such avatars may last from several months to a few years. 
Fortunately, the attack target and the required number of avatars are usually not known in advance significantly reducing the cost-effectiveness of sophisticated avatars. 

\emph{Chameleon} profiles exposed in this article facilitate efficient management of a fleet of avatars by maintaining a pool of mature avatars whose timeline can be adapted to the agenda of the attack target once it is known.   

\paragraph{Execution Flow}
In this special misuse case the attack phases  \emph{weaponizing} and \emph{maturation} are performed twice; both before and after the attack target is known. 

\noindent(1) The first \emph{weaponizing} phase starts with establishing the redirect infrastructure and building a fleet of avatars. 
They are created with neutral displays common within the OSN.  

\noindent(2) During the initial \emph{maturation} process the neutral avatars regularly publish \chameleon posts with neutral displays. 
They acquire friends within the OSN while maximizing the acceptance rate of their friend requests~\cite{paradise2017creation}.

\noindent (3) Once the attack target is known the attacker performs the required \emph{reconnaissance}, selects some of the mature \chameleon profiles, and (4) \emph{weaponizes} them with the relevant agenda by changing the profile information and the display of all past \chameleon posts.

\noindent During (5) \emph{delivery} and (6) the second \emph{maturation} phase the refreshed \chameleon profiles (avatars) contact the target and build trust with it.    
The (7) \emph{execution} phase in this misuse case depends on the attacker's goals. 
It is likely that the avatars that already engaged in an attack will be discarded.

\paragraph{Potential Impact}
It's unnecessary for the adversary to create an OSN account and build an appropriate agenda for each avatar long before executing an attack.  
\emph{Chameleon} profiles and posts are created and maintained as a general resource suitable for various attack campaigns.  
As a result, the cost of maintaining such avatars is dramatically reduced. 
Moreover, if an avatar is detected and blocked during the attack campaign, its replacement can be \emph{weaponized} and released very quickly.

\subsubsection{Evading Censorship} 
OSNs maintain millions of entities, such as pages, groups, communities, etc.
For example, Facebook groups unite users based on shared interests~\cite{casteleyn2009use}.
In order to ensure proper language, avoid trolling and abuse, or allow in only users with a very specific agenda, moderators inspect the users who ask to join the groups and review the published posts. 
\emph{Chameleon} attack can help in evading censorship, as well as a shallow screening of OSN profiles.   
See Section~\ref{sec:mitigation} for specific recommendations on profile screening to detect \chameleon profiles.

For example, assume two Facebook groups, uniting Democrat and Republican activists during USA elections. 
Assume a dishonest activist from one political extreme that would like to join a Facebook group of the rivals. 
Reasons may vary from trolling to spying. 
Assume, that this activist would like to spread propaganda within the rival group.   
But pages that exhibit an agenda that is not appropriate for the group would not be allowed by the group owner. 
The next procedure would allow the rival activist to bypass the censorship of the group moderator.

\paragraph{Execution Flow}
\noindent During the (1) \emph{reconnaissance} phase, the adversary learns the censorship rules of the target.  
\noindent(2) The \emph{weaponizing} phase includes establishing a \chameleon profile with agenda appropriate to the censorship.
During the \noindent(3) \emph{maturation} phase, the adversary publishes posts with redirect links to videos fitting the censorship rules. 
(4) \emph{delivery} in this case represents the censored act such as requesting to enter a group, sending a friend request, posting a video, etc. 
The censor, e.g. the group's administrator, reviews the profile and its timeline and approves them to be presented to all group members. 
Finally, in the \noindent(5) \emph{execution} phase, the adversary changes the display of its profile and posts to reflect a new agenda that would otherwise not be allowed by the censor.

\paragraph{Potential Impact}
This attack allows the adversary to infiltrate a closed group and publishing posts in contrast to the administrator's policy. 
Moreover, one-time censorship of published content would no longer be sufficient. 
Moderators would have to invest a lot more effort in the periodical monitoring of group members and their posts to ensure that they still fit the group's agenda.  
In Section~\ref{sec:group_infiltration_experiments}, we demonstrate the execution of the \chameleon attack for penetrating closed groups using soccer fan groups as an allegory for groups with extreme political agenda.

\subsubsection{Promotion} 

Unfortunately, the promotion of content, products, ideas, etc. using bogus and unfair methods is very common in OSNs. 
Spam and crowdturfing are two example techniques used for promotion. 
The objective of spam is to reach maximal exposure through unsolicited messages. 
Bots and crowdturfers are used to misrepresent the promoted content as a generally popular one by adding likes and comments. 
Crowdturfers~\cite{lee2013crowdturfers} are human workers who promote social content for economic incentives.
\emph{Chameleon} attack can be used to acquire likes and comments of genuine OSN users by piggybacking a popular content.

\paragraph{Execution Flow}

During \noindent(1) \emph{reconnaissance} phase, the attacker collects information about a topic favorable to the general public that is related to the unpopular content that the attacker wants to promote. \noindent(2) During the \emph{weaponizing} phase, the attacker creates \chameleon page and posts that support the favorite topic.
For example, assume an adversary who is a new singer who would like to promote themselves. 
In the \emph{weaponizing} phase, he/she can create a \chameleon page that supports a well-known singer.
During the \noindent(3) \emph{delivery} and \noindent(4) \emph{maturation} phases, the OSN users show their affection towards seemingly appealing content by interacting with the \chameleon page using likes, linking, retweeting, commenting, etc. 
As time passes, the \chameleon page obtains social capital.  
In the final (5) \emph{execution} phase, the \chameleon page's display changes to support the artificially promoted content retaining the unfairly collected social capital.

\paragraph{Potential Impact}
The attacker can use \chameleon pages and posts to promote content in OSNs by piggybacking popular content.  
The attacker enjoys the social capital provided by a genuine crowd that otherwise would not interact with the dubious content.
Social capital obtained from bots or crowdturfers can be down-rated using various reputation management techniques. 
In contrast, social capital obtained through the \chameleon trickery is provided by genuine OSN users.

\subsubsection{Clickbait}

Most of the revenues of online media come from online advertisements~\cite{chakraborty2016stop}.
This phenomenon generated a significant amount of competition among online media websites for the readers' attention and their clicks.
To attract users and encourage them to visit a website and click a given link, the website administrators use catchy headlines along with the provided links, which lure users into clicking on the given link~\cite{chakraborty2016stop}.
This phenomenon titled clickbait.

\paragraph{Execution Flow}

\noindent(1) During the \emph{weaponizing} phase, the attacker creates \chameleon profiles with posts with redirect links. Consider an adversary that is a news provider who would like to increase the traffic to its website.
To increase its revenues, he can do the following:
in the \emph{weaponizing} phase, he should publish a \chameleon post with a catchy headline with an attached link to an interesting article.
Later, in the \emph{maturation} phase, users attract the post by its attractive link preview, as well as its headline, and increase the traffic to a website.
Later, in the  \emph{execution} phase,  the adversary changes the redirect target of the posted link but keeping the link preview not updated.
As a result, new users will click on the \chameleon post that its display did not change, but now they will be navigated to the adversary's website.

\paragraph{Potential Impact}
By applying this attack, the attacker can increase his traffic and, eventually, his income. Luring the users with an attractive link preview in which increases the likelihood that the user will click on it and will consume his content.

\section{Susceptibility of Social Networks to the Chameleon Attack}
\label{sec:susceptibility}
 
\subsection{Online Social Networks}
We review the susceptibility of seven OSNs to the \chameleon attack. 

\begin{table}[ht]
\caption{Features that facilitate and mitigate the \chameleon attack in popular OSNs. } 
\small
\begin{tabular}{|p{2.5cm}|l|l|l|l|l|l|l|l|}
\hline
   Attacker's ability  &OSN feature &Facebook &Twitter &WhatsApp &Instagram &Reddit &Flickr &LinkedIn \\
  \hline
  \hline
\multirow{2}{2.5cm}{Creating artificial timeline}  
&Editing a post's publication date & Y \faThumbsDown & N \faThumbsOUp & N \faThumbsOUp & N \faThumbsOUp & N \faThumbsOUp & N \faThumbsOUp & N \faThumbsOUp  \\
   \cline{2-9}
&Presenting original publication date & Y \faThumbsOUp & - & - & - & - & - & -  \\  
  \hline
  \hline
\multirow{3}{*}{Changing content} &Editing previously published posts & Y \faThumbsDown & N \faThumbsOUp & N \faThumbsOUp & Y \faThumbsDown & Y \faThumbsDown & Y \faThumbsDown & Y \faThumbsDown \\
   \cline{2-9}
&Indication in edited published posts  & Y \faThumbsOUp & - & - & Y \faThumbsOUp & N \faThumbsDown & N \faThumbsDown & Y \faThumbsOUp \\
   \cline{2-9}
&Indication in edited shared posts & N \faThumbsDown & - & - & Y \faThumbsOUp & - & - & Y \faThumbsOUp
   \\
\cline{2-9}
&Presenting edit history  & Y \faThumbsOUp & - & - & N \faThumbsDown & - & - & N \faThumbsDown \\
   \hline
   \hline
\rowcolor{lightgray} & Publishing redirect links & Y {\faThumbsDown}  & Y \faThumbsDown & Y \faThumbsDown & N \faThumbsOUp & Y \faThumbsDown & Y \faThumbsDown & Y \faThumbsDown \\
   \cline{2-9}
\rowcolor{lightgray} Changing display &Displaying link preview & Y \faThumbsDown & Y \faThumbsDown & Y \faThumbsDown & - & Y \faThumbsDown & N \faThumbsOUp & Y \faThumbsDown \\
   \cline{2-9}
\rowcolor{lightgray}&Updating link preview & Y \faThumbsDown & Y \faThumbsDown & N \faThumbsOUp & - & N \faThumbsOUp & - & Y \faThumbsDown \\
   \hline
   \hline
Switching content &Hiding posts & Y \faThumbsDown & N \faThumbsOUp & N \faThumbsOUp & Y \faThumbsDown & Y \faThumbsDown & Y \faThumbsDown & N \faThumbsOUp  \\
   \hline
\end{tabular}
\faThumbsDown = Facilitates the \chameleon attack.   \faThumbsOUp = Mitigates the \chameleon attack
\label{tab:OsnCompareTbl}
\end{table}

\subsubsection{Facebook}
Alongside its growing popularity, Facebook allows users to manipulate the display of previously published posts based on several different features.  
The features include the publishing of redirect links, editing post's publication date, hiding previously published posts, and publishing unauthorized content in a closed group.

Up until 2017, in case a user edits a post on Facebook, an indicator is presented for the users to notify them that the content had been updated. 
After 2017, a Facebook update removed this public notification and enable to see the post's history only via the button of 'View Edit History.'

While Facebook allows editing post's publication date, it displays a small indication concerning the original publication date of the post. 
To watch the original publication date, a user must hover over the clock icon shown in the post, and a bubble will be shown together with the original publication date. 

Also, concerning Facebook pages, Facebook does not allow to do radical changes to the original name of a page daily. 
However, it is still possible to conduct limited edits to the page's name; changes that are so minor that the context of the original name will be not changed. 
As a result, we were able to rename a page in phases by editing the name of a given page with small changes in each edit action. 
First, we changed only two characters from the name of a page. 
Afterward, three days later, we changed two more characters and so forth until eventually, we were able to rename the page entirely as we wished.

Facebook employs a mechanism called Link Shim to keep Facebook users safe from external malicious links \cite{FacebookLinkShim}. 
When clicking on an external link posted on Facebook, their mechanism checks if the link is  blacklisted. 
In case of a suspicious URL, Facebook will display a message to the user \cite{FacebookLinkShim}. 
Redirect links used in \chameleon posts lead to legitimate destinations and so are currently approved by Link Shim.

\subsubsection{Twitter}
As opposed to Facebook, Twitter does not allow users to edit and hide tweets that have already been published, or to manipulate a tweet's publication date (see Table \ref{tab:OsnCompareTbl}). 
This mechanism makes it more difficult for an attacker to manipulate the display of previously published content.
On the other hand, Twitter allows the use of client redirects. 
This poses the same danger as Facebook redirects, allowing attackers to manipulate the link preview of a tweet with content that is not necessarily related to the target website. 
Moreover, Twitter allows users to update a link preview using the \emph{Card Validator}.\footnote{ \url{https://cards-dev.twitter.com/validator}}
In addition, Twitter makes it possible to change a user's display name but does not allow to change the original username chosen during registration (serves as an identifier).

\subsubsection{WhatsApp}
WhatsApp allows messages to be published with redirect links and it displays link previews but it does not allow the update of an already published link preview.  
As opposed to other OSNs, WhatsApp is the only OSN that displays an indication that the message was deleted by its author. 

WhatsApp is safe against most flavors of the \chameleon attack, except \emph{clickbait} where an attacker can trick other users by encouraging them to click a malicious link with a preview of a benign link.

\subsubsection{Instagram}
Concerning redirect links, Instagram does not allow users to publish external links (see Table~\ref{tab:OsnCompareTbl}). 
Since the posts on Instagram are based on images, it is not an option for the attacker to change the published content by redirect link. 
However, Instagram allows to edit already published posts, The editing process includes the text in the description section, as well as the image itself.
In case of such a change to a post was made by its owner, no indication is shown to users.

\subsubsection{Reddit}
Alongside its popularity, Reddit is prone to a \chameleon attack: In this OSN, the attacker can edit, delete or hide already published posts while others will not be able to know that the content has been modified.

\subsubsection{Flickr}
As opposed to Facebook and WhatsApp, Flickr does not show link previews, but it allows users to update their posts, replace uploaded images, hide already published posts, and edit their account name.
All these activities can be performed by users, without any indication for the users to the editing activity.

\subsubsection{LinkedIn}
LinkedIn permits users to share a redirect link, and to update the link preview using \emph{Post Inspector}.\footnote{ \url{https://www.linkedin.com/post-inspector}}
As a result, users can edit their posts, however, the post will be marked as edited.

\subsection{Existing Weaknesses and Security Controls}
\label{sec:weaknesses}
Next, we summarize the OSN weaknesses related to the \chameleon attack, as well as controls deployed by the various OSNs to mitigate potential misuse. 
While the main focus of this article is the \chameleon attack facilitated by cached link previews, in this subsection we also discuss other types of the \chameleon attack successfully mitigated by major OSNs.   

\subsubsection{Creating artificial timeline}
Publishing posts in retrospective is a feature that is easiest to misuse.  
Such a feature helps an adversary creating OSN accounts that look older and more reliable than they are. 
Luckily, all OSNs, but Facebook, do not provide such a feature to their users. 
Although Facebook allows \emph{editing a post's publication date}, it mitigates possible misuse of this feature for creating artificial timelines by \emph{presenting the original publication date} of the post.

\subsubsection{Changing content}
Some OSNs provide their users with the ability to \emph{edit previously published posts}.   
This feature facilitates all misuse cases detailed in Section~\ref{sec:misuse} without any additional resources required from the attacker.  
Twitter and WhatsApp do not allow editing of previously published posts. 
Facebook, Instagram, and LinkedIn mitigate potential misuse by \emph{presenting editing indication in published posts}. 
Facebook even \emph{presents the edit history} of a post. 
Unfortunately, in contrast to Instagram and LinkedIn, Facebook does not \emph{present the edit indication in shared posts}. 
We urge Facebook to correct this minor yet important omission.

\subsubsection{Changing Display}
The primary weakness of the major OSNs (Twitter, Facebook, and LinkedIn) which facilitates the \chameleon attack discussed in this paper is the combination of three features provided by the OSNs. 
First, \emph{publishing redirect links} allows attackers to change the navigation target of the posted links without any indication of such a change.
Second, OSNs \emph{display a link preview} based on metadata provided by the website at the end of the chain of server redirects.  
This feature allows the attackers to control the way link previews are displayed.
Finally, OSNs allow \emph{updating link preview} following the changes in the redirect chain of the previously posted link. 
Such an update is performed without displaying an indication that the post was updated.
Currently, there are no controls that mitigate the misuse of these features. 

WhatsApp, Reddit, and LinkedIn display link previews of redirect links similar to Facebook and Twitter. 
But they do not provide a feature to update the link previews. 
On one hand, the only applicable misuse case for the \chameleon attack in these OSNs is \emph{clickbait}. 
On the other hand, updating link previews is important for commercial brand management.  

\subsubsection{Switching Content}
Facebook, Instagram, Reddit, and Flickr allow users to temporarily \emph{hide their posts}. 
This feature allows a user to prepare multiple sets of posts where each set exhibits a different agenda. 
Later, the adversary may display the appropriate set of posts and hide the rest.
The major downsides of this technique as far as the attacker is concerned are: (1) The need to maintain the sets of posts ahead of time similar to maintaining a set of regular profiles. (2) Social capital acquired by one set of posts cannot be reused by the other sets, except friends and followers.     

Overall all reviewed OSNs are well protected against timeline manipulation. 
The major OSNs, except Reddit and Flickr, are aware of the dangers of post-editing and provide appropriate controls to avoid misuse. 
Due to the real-time nature of messaging in Twitter and WhatsApp, these OSNs can disable the option of editing posts.

The major OSNs, Facebook, Twitter, and LinkedIn, really care about the business of their clients and thus, explicitly provide features to update link previews.  
\emph{Chameleon} attack exposed in this paper misuses this feature to manipulate the display of posts and profiles. 
Provided that Reddit and Flickr allow editing the post content, only WhatsApp and Instagram are not susceptible to \chameleon attacks.

Instagram stores the posted images and not the links to the external resources, an approach that may not scale and may not be suitable for all premium customers.  
WhatsApp stores the data locally and does not allow recollecting past messages if the receiver is not a member of the group when the message was posted.  
Clearly, WhatsApp's approach is not suitable for bloggers, commercial pages, etc. that would like to share their portfolio with every newcomer.

\subsection{Additional Required Security Controls}
\label{sec:mitigation}

The best way to mitigate the \chameleon attack is to disallow redirect links and to disable link preview updates in all OSNs. 
Nevertheless, we acknowledge that it is not possible to stop using external redirect links and short URLs. 
These features are very popular on social networks and important in brand management.

First and foremost an appropriate change indication should be displayed whenever the link preview cache is updated. 
Since on Facebook the cache is updated by the author of the original post, this act can naturally be displayed in the edit history of the post.  
Link preview cache updates should be treated similar to the editing of posts.

However, edit indications on posts won't help unless OSN users will be trained to pay attention to these indications. 
Facebook, and other OSNs, should make it crystal clear which version of the post a user liked or commented on.  
To minimize the impact of the \chameleon attack likes, shares and comments of a post should be associated with a specific version of the post within the edit history, by default. 
It is also important to let users know about subsequent modifications of the posts they liked, commented, or shared.  
The users will be able, for example, to delete their comments or to confirm it, moving the comment back from the history to the main view.  

In Twitter and LinkedIn, anyone can update the link preview. 
The motivation for this feature is two-fold:
(1) The business owner should be able to control the look and feel of his business card within the OSN regardless of the specific user who posted it. 
(2) Link previews should always be up to date. 
It will be challenging to design appropriate mitigation for the \chameleon without partially giving up these objectives. 

We suggest notifying a Twitter (or LinkedIn) user who posted a link to an external site whenever the link preview is updated. 
The user will be able to delete the post or accept the link preview update at his sole discretion. 
By default, the link preview should remain unchanged.  
This approach may increase the number of notifications the users receive, but with appropriate filters, it will not be a burden on the users. 
However, it may require maintaining copies of link previews for all re-posted links, which in turn significantly increase storage requirements.  

Finally, OSNs should update their anomaly detection algorithms to take into account changes made to the posts' content and link previews as well as the reputation of the servers along the redirection path of the posted links.

It may take time for the OSNs to implement the measures described above. 
Meanwhile, users should be aware that their \emph{likes and comments are precious assets} that may be used against them if given out blindly. 

Next, we suggest a few guidelines that will help average OSN users detecting \chameleon posts and profiles. 
Given a suspected profile, check the textual content of its posts.   
\emph{Chameleon} profiles should publish general textual descriptions to easily switch agenda.  
The absence of opinionated textual descriptions in the topic of your mutual interest may indicate potential \chameleon. 
A Large number of ambiguous posts that can be interpreted in the context of the cover image or in the context of other posts in the timeline should increase the suspicion. 
For example, ``This is the best goalkeeper in the world!!!'' without a name mentioned is ambiguous. 
Also using public services like Facebook provided\footnote{ https://developers.facebook.com/tools/debug/sharing/batch/} for watching a given post history can be useful for detecting a \chameleon post.

A large number of redirect links within the profile timeline is also an indication of \chameleon capabilities. 
We do not encourage the users to click links in the posts of suspicious profiles to check whether they are redirected! 
In Facebook and LinkedIn, a simple inspection of the URL can reveal whether a redirection is involved. 
Right-click the post and copy-paste the link address in any URL decoder. 
If the domain name within the copied URL matches the domain name within the link preview and you trust this domain, you are safe. 
Today, most links on Facebook are redirected through Facebook's referral service. 
The URL you should look at follows the ``u'' parameter within the query string of l.facebook.com/l.php. 
If the domain name is appearing after ``, u='' differs from the domain name within the link preview, the post author uses redirection services. 
Unfortunately, today, links posted on Twitter are shortened, and the second hop of the redirection cannot be inspected by just copying the URL.

\section{Group Infiltration Experiment}
\label{sec:group_infiltration_experiments}
In this section, we present an experiment conducted on Facebook to asses the reaction of Facebook group moderators to \chameleon pages. 
In this experiment, we follow the execution flow of the misuse case number 4 \emph{evading censorship} in Section~\ref{sec:misuse}.

\subsection{Experimental Setup}
\label{sec:experimentalSetup}

In this experiment, four pairs of rival soccer and basketball teams were selected:
Arsenal vs Chelsea, Manchester United vs Manchester City, Lakers vs Clippers, and Knicks vs Nets.
We used sixteen Facebook pages: one regular and one \chameleon page for each sports teams.  
Regular pages post YouTube videos that support the respective sports team. 
Their names are explicitly related to the team they support e.g. ``Arsenal - The Best Team in the World.'' 
\emph{Chameleon} pages post redirect links that lead to YouTube videos that support either the team or their rivals. 
Their names can be interpreted based on context e.g. ``The Best Team in the World.''  
The icons and cover images of all pages reflect the team they (currently) support.

Next, we selected twelve Facebook groups that support each one of the eight teams (total of 96 Facebook groups) according to the following three criteria: (a) the group allows pages to join it, (b) the group is sufficiently large (at least 50 members), and (c) there was at least some activity within the group in last month.  

We requested to join each group four times: 
(1) as a regular fan page, (2) as a regular rival page, (3) as a \chameleon page while supporting the rivals, and (4) the same \chameleon page requested to join the group again now pretending to be a fan page. We requested each group in a random order of the pages.
We used balanced experiment design to test all permutations of pages where the respective \chameleon page first requests to join the group as rival's page and afterward as fan's page. 
We allowed at least five days between consequent requests to join each group.

A page can be \emph{Approved} by the group admin or moderator (hereafter admin). 
In this case the page becomes a member of the group. 
While the admin have not decided yet, the request is \emph{Pending}.
The owner can \emph{Decline} the request. 
In this case the page is not a member of the group, but it is possible to request to join the group again. 
Neither one of our pages was \emph{Blocked} by the group admins, therefore, we ignore this status in the following results. 
Whenever, \chameleon page pretending to be a rival page is \emph{Approved} by an admin, there is no point in trying to join the same group using the same page again. 
We consider this status as \emph{Auto Approved}.

The first phase of the experiment started on July 20, 2019 and included only the Facebook groups supporting Chelsea and Arsenal.
The relevant \chameleon pages changed the way they are displayed on Aug. 16.
The second phase started on Sept. 5, 2019 and included the rest of the Facebook groups. 
The relevant \chameleon pages changed the way they are displayed on Sept. 23. 
The following results summarize both phases.

\subsection{Results}
\label{sec:results}

\begin{figure}[t]
\centering
\includegraphics[width=3.5in]{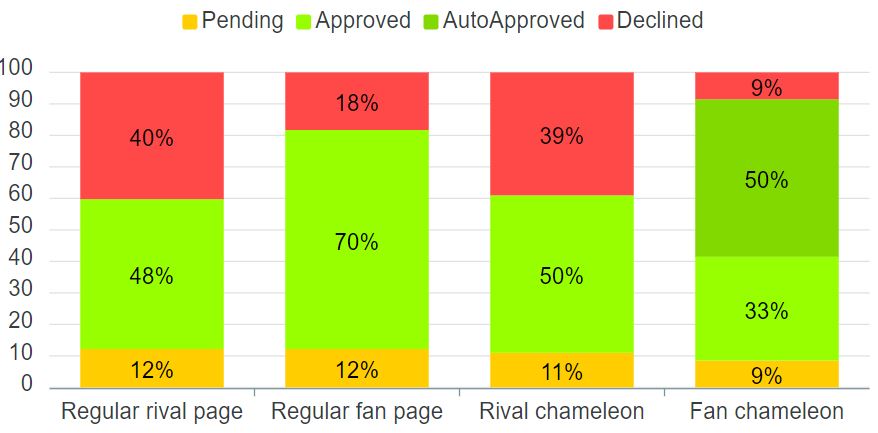}
\caption{Request results by type of page}
\label{fig:chamVsPage}
\end{figure}

During the experiment, 14 Facebook groups decided to prevent (any) pages from joining the group. 
We speculate that the admins were not aware to the option of accepting pages as group members, and updated the group settings after they saw our first requests.  
These 14 groups were \emph{Disqualified} in current experiment. 
Overall there were 206 \emph{Approved} requests, 87 \emph{Declined}, and 35 \emph{Pending}.
Figure~\ref{fig:chamVsPage} presents the distribution of request statuses for the different types of pages.

Some admins blindly approve requests. 
For example, 28 groups approved all requests. 
Other group admins meticulously check the membership requests.   
Thirteen groups \emph{Declined} or ignored the rival pages and \emph{Approved} pages that exhibit the correct agenda.

Overall, \textbf{the reaction of admins to \chameleon pages is similar to their  reaction to regular pages with the same agenda}. 
To check this hypothesis we used a one-way ANOVA test to determine whether there is a significant difference between the four types of group membership requests.
The test was conducted on the request status values at the end of the experiment (\emph{Declined}, \emph{Pending}, \emph{Approved}). 
Results showed that there is no statistically-significant difference between the approval of \chameleon fan pages and regular fan pages (p-value = 0.33). 
There is also no statistically-significant difference between the approval of \chameleon rival pages and regular rival pages (p-value = 0.992). 
However, the difference between approval of either regular or \chameleon rival pages and the approval of both types of fan pages is statistically-significant with p-value ranging from 0.00
to 0.003. 
These results indicate that the reaction of admins to \chameleon pages in our experiment is similar to their reaction to regular (non-\emph{chameleon}) pages with a similar agenda.  
We conclude that \textbf{admins do not distinguish between regular and \chameleon pages.} 
This conclusion is stressed by the observation that only two groups out of 82 \emph{Declined} \chameleon fan pages and \emph{Approved} regular fan pages. 
Seven groups approved \chameleon fan pages and rejected regular fan page.

The above results also indicate that, in general, \textbf{admins are selective toward pages that they censor.} 
Next, we measure the selectivity of the group admins using Likert scale~\cite{joshi2015likert}.
Relying on the conclusion that admins do not distinguish between regular and \chameleon pages, we treat them alike to measure admins' selectivity. 
Each time a group admin \emph{Declined} a rival page or \emph{Approved} a fan page he/she received one point. 
Each time a fan page was \emph{Declined} or a rival page was \emph{Approved}, the selectivity was reduced by one point. 
\emph{Pending} request status added zero toward the selectivity score.

For each group, we summed up the points to calculate its selectivity score. 
When selectivity score is greater than three, we consider the group as \emph{selective}. 
If the selectivity score is less than or equal to three, we consider the group as \emph{not selective}.

\begin{figure}[h]
\centering
\includegraphics[width=3.5in]{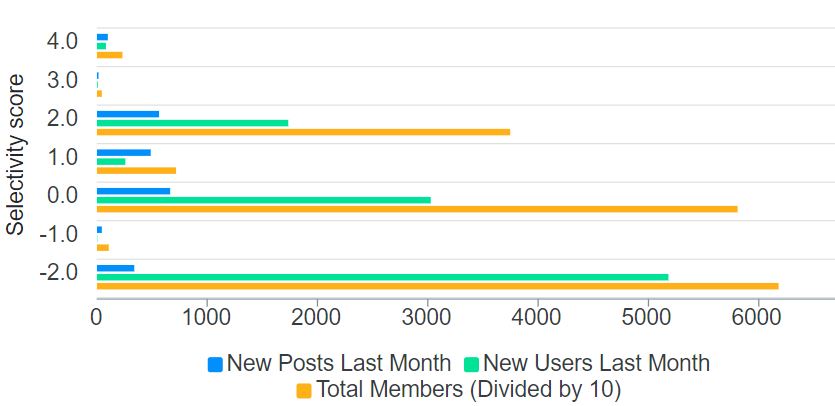}
\caption{Average groups activity by selectivity score}
\label{fig:groupsScoreAvg}
\end{figure}

To explain the differences in groups' selectivity we first tested whether there is a difference between the number of members in selective and non-selective groups using t-tests. 
We found that smaller groups are more selective than, larger ones with (p-value = 0.00029). 
This result is quite intuitive. 
Smaller groups tend to check the identity of the users who ask to join the group, while large groups are less likely to examine the identity of the users who want to join the group. 
Figure~\ref{fig:groupsScoreAvg} presents the  groups' activity and size vs. their selectivity score. 
There is a weak negative correlation between group's selectivity score and number of members (Pearson correlation = -0.187, p-value = 0.093).

\section{Related Work}
\subsection{Content Spoofing and Spoofing Identification} 
Content spoofing, also known as content injection or virtual defacement, is an attack method that deceives users in assuming that particular content on a web site is legitimate and not from an external source \cite{lungu2010optimizing, awang2013detecting, karandel2016security}. Using this attack, an attacker can upload new, fake, or modified content to the web site as legitimate. 
Content spoofing can lead to malware exposure, financial fraud, or privacy violations, and can misrepresent an organization or an individual \cite{hayati2009spammer, benea2012anti}. Content spoofing usually exploits the trust relationship between a web application and its users as the modified web page is presented to the user under the context of the application domain \cite{hussain2019content}.

According to WhiteHat \cite{grossman2017whitehat}, content spoofing is one of the most prevalent vulnerabilities in web applications. 
The content spoofing attack leverages the code injection vulnerability where the user's input is not sanitized correctly. 
Using this vulnerability, an attacker can provide new content to the web, usually via the GET or POST parameter. 
Hussain et al. \cite{hussain2019content} present two ways to conduct content spoofing attack. 
An HTML Injection, in which the attacker alters the content of a web page for malicious purposes by using HTML tags, or a Text Injection that manipulates the text data of a parameter. 

Jitpukdebodin et al. \cite{jitpukdebodin2014novel} introduce a new technique that explores vulnerability in WLAN communication. The proposed method creates a crafting spoof web content and sends it to a user before the genuine web content from a website is transmitted to the user. 
Hussain et al. \cite{hussain2019content} present a new form of compounded SQL injection attack technique which uses the SQLi attack vectors to perform content spoofing attacks on a web application.

There have been a few techniques for the detection of content spoofing attacks \cite{benea2012anti, niemela2018detecting}. 
Benea et al. \cite{benea2012anti} suggest a system to prevent content spoofing by detecting phishing attacks using fingerprints similarity. 
Niemela and Kesti \cite{niemela2018detecting} present a method for detecting unauthorized changes to a website using authorized content policy sets for each of a multiplicity of websites from the web operators and identify websites that do not conform to respective policy sets.

\subsection{Website Defacement}
Website Defacement is an attack that changes the visual appearance of a website \cite{kanti2011implementing, borgolte2015meerkat, romagna2017hacktivism}. Using this attack, an attacker can cause serious consequences to website owners, including interrupting website operations and damaging the owner's reputation. More interestingly, attackers may support their reputation, promoting a certain ideological, religious, or political orientation \cite{romagna2017hacktivism,maggi2018investigating}. Besides, web defacement is a significant threat to businesses since it can detrimentally affect the credibility and reputation of the organization \cite{borgolte2015meerkat,medvet2007detection}. Most website defacement occurs when attackers manage to find any vulnerability in the Web application and then inject a remote scripting file \cite{kanti2011implementing}.

Several types of research deal with the monitoring and detection of website defacement, with solutions that include signature-based detection \cite{gurjwar2013approach, shani2010system} and anomaly-based detection \cite{borgolte2015meerkat,davanzo2011anomaly,hoang2018website}.

The simplest method to detect website defacement is a checksum comparison. With this method, the content of a website is calculated using hashing algorithms. The website is then monitored and a new checksum is calculated and compared with the previous one, raising an alert when the checksum is changed \cite{kanti2011implementing, gurjwar2013approach, shani2010system}. 
This method is effective for static web pages but not applicable to dynamic pages.

Several techniques have been proposed for website defacement based on complex algorithms \cite{medvet2007detection,kim2006advanced}. Kim et al. \cite{kim2006advanced} propose using a 2-gram method for building a profile from normal web pages for monitoring and detecting of page defacement. Medvet et al. \cite{medvet2007detection} present an approach to automatic detection of website defacement based on genetic programming. The method does not rely on any domain-specific knowledge but instead builds an algorithm based on a sequence of readings of the remote page to be monitored, and on a sample set of attacks. The remote page is then monitored at regular intervals; the algorithm is applied, which raises an alert when a suspect modification is found.

Several techniques use machine learning-based methods for website defacement detection \cite{borgolte2015meerkat,davanzo2011anomaly,hoang2018website,bartoli2006automatic}. Those studies, build a profile of the monitored page automatically, based on machine learning techniques, and raise an alert when the page content does not fit the profile. Borgolte et al. \cite{borgolte2015meerkat} proposes the 'MEERKAT' detection system that requires no prior knowledge about the website content or its structure, but only its URL. 'MEERKAT' automatically learns high-level features from screenshots (image data) of defaced websites by machine learning, such as stacked autoencoders and deep neural networks. Results on the largest website defacement dataset achieve a high detection accuracy of over 97\% and a low false-positive rate of less than 1.5\%. The method's drawback is that it requires extensive computational resources for image processing and recognition. Recently advanced research \cite{bergadano2019defacement} proposes an application of adversarial learning to defacement detection for making the learning process unpredictable so that the adversary will be unable to replicate it and predict the classifier's behavior using a secret key.

\subsection{Cloaking attack and identification}
Cloaking, also known as 'bait and switch' is a common technique used to hide the true nature of a Web site by delivering different semantic content to some selected specific user group-based \cite{wang2011cloak, invernizzi2016cloak}. Wang and Savage \cite{wang2011cloak} presented four cloaking types: Repeat cloaking (delivering different web content based on visit times of visitors), User-agent cloaking (delivering specific web content based on visitors' User-agent String), Redirection cloaking (redirecting users to another website by using JavaScript), and IP Cloaking (delivering specific web content based on visitors' IP).

Researchers have responded to the cloaking techniques with a variety of anti-cloaking techniques \cite{invernizzi2016cloak}. Basic techniques relied on a cross-view comparison technique \cite{wang2006detecting, wang2007spam}. A page is classified as cloaking if the redirect chain deviated across fetches. Other approaches mainly target compromised webservers and identify clusters of URLs with trending keywords that are irrelevant to the other content hosted on page \cite{john2011deseo}. Wang et al. \cite{wang2011cloak} identify cloaking in near real-time by examining the dynamics of cloaking over time. Invernizzi et al. \cite{invernizzi2016cloak} develop an anti-cloaking system that detects split-view content returned to two or more distinct browsing profiles by building a classifier that detects deviations in the content.

\subsection{Manipulating Human Behavior} 

These days, cyber-attacks manipulate human weaknesses more than ever \cite{blunden2010manufactured}. Our susceptibility to deception, an essential human vulnerability, is a significant cause of security breaches. 
Attackers can exploit the human vulnerability by sending a specially crafted malicious email, tricking humans into clicking on malicious links, and thus downloading malware, (a.k.a. spear phishing) \cite{goel2017got}.

One of the main attack tools that exploit the human factor is social engineering, which is defined as the manipulation of the human aspect of technology using deception \cite{uebelacker2014social}. Social engineering plays on emotions such as fear, curiosity, excitement, and empathy, and exploits cognitive biases \cite{abraham2010overview}. The basic 'good' human nature characteristics make people vulnerable to the techniques used by social engineers, as it activates various psychological vulnerabilities, which could be used to manipulate the individual to disclose the requested information \cite{bezuidenhout2010social, conteh2016cybersecurity, conteh2016rise, luo2011social} 

The exploitation of the human factor has extensive use in advanced persistent threats (APTs). An APT attack involves sophisticated and well-resourced adversaries targeting specific information in high-profile companies and governments \cite{chen2014study}. In APT attacks, social engineering techniques are aimed at manipulating humans into delivering confidential information about a targeted organization or getting an employee to take a particular action (e.g., executing malware) \cite{paradise2017creation, gulati2003threat, bere2015advanced}.

To the best of our knowledge, the \chameleon attack was previously executed in files during content-sniffing XSS attacks \cite{barth2009secure} but not on the social media platform. Chameleon documents discussed by Barth et al. are files conforming to multiple file formats (e.g. PostScript+HTML). The attack exploits the fact that browsers can parse documents as HTML and execute any hidden script within. In contrast to chameleon documents, which are parsed differently by different tools without adversarial trigger, our chameleon posts are controlled by the attacker and are presented differently to the same users at different times.

Lately, Stivala and Pellegrino~\cite{Stivala2020deceptive} conducted a study associated with link previews independently.
In their research, they analyzed the elements of the preview links during the rendering process within 20 OSNs and demonstrated a misuse case by crafting benign-looking link previews that led to malicious web pages.

\section{Conclusions and Future Work}
This article discloses a weakness in an important feature provided by three major OSNs: Facebook, Twitter, and LinkedIn, namely \emph{updating link previews without visible notifications while retaining social capital} (e.g., likes, comments, retweets, etc.). 
This weakness facilitates a new \chameleon attack where the link preview update can be misused to damage the good name of users, avoid censorship, and perform additional social networks scam detailed in Section~\ref{sec:misuse}.   
Out of seven reviewed social networks, only Instagram and WhatsApp are resilient against most flavors of the \chameleon attack. 

We acknowledge the importance of the link preview update feature provided by the OSNs to support businesses that disseminate information through social networks 
and suggest several measures that should be applied by the OSNs to reduce the impact of \chameleon attacks.  
The most important measure is binding social capital to the version of a post to which it was explicitly provided.  
We also instruct OSN users on how to identify possible chameleons.    

We experimentally show that even the most meticulous Facebook group owners fail to identify \chameleon pages trying to infiltrate their groups. 
Thus it is extremely important to raise the awareness of OSN users to this new kind of trickery. 

We encourage researchers and practitioners to identify potential \chameleon profiles throughout the OSNs in the nearest future; 
develop and incorporate redirect reputation mechanisms into machine learning methods for identifying social network misuse;
and include the \chameleon attack in security awareness programs alongside phishing scam and related scam.

\section{Ethical and Legal Considerations}
\label{sec:ethics}
Our goal is hardening OSNs against misuse while respecting the needs and privacy of OSN users. 
We believe that it is important to raise the awareness of researchers, practitioners, OSN operators and users to the potential misuse of link previews.   
We follow strict Responsible Full Disclosure Policy, as well as guidelines recommended by the Ben-Gurion University Human Subject Research Committee. 

In particular, we did not access or store any information about the profiles we contacted during the experiment. 
We only recorded the status of the requests to join their Facebook groups. 
The \chameleon pages used during the experiment are deleted at the end of the study.
Owners of the contacted Facebook groups are able to decide whether or not to accept the request from our pages. 
Although we did not inform them about the study prior to the requests, they are provided with post-experiment written feedback regarding their participation in the trial, as well as guidelines to safeguard themselves from \chameleon attacks and other OSN scam.
We contact the relevant OSNs at least one month prior to publication of the trial results and disclosure of the related weaknesses.
No rules or agreements were violated in the process of this study.   
In particular, we use Facebook pages in the showcase and in the experiment rather than profiles in order to adhere to the Facebook End-User Licence Agreement. 




\section{Availability}
   
\emph{Chameleon} pages, posts, and tweets are publicly available.
Links can be found in the following GitHub repository:
\url{https://github.com/aviade5/Chameleon-Attack/}.

Source code is not provided to reduce misuse. 
CWE and official responses of the major OSNs are also provided on the mentioned GitHub page.

\newpage\clearpage

\bibliographystyle{unsrt}
\bibliography{draft_for_arxiv.bib}
\newpage

\end{document}